# Genetic influences on translation in yeast

Frank W. Albert[1], Dale Muzzey[2], Jonathan Weissman[2], and Leonid Kruglyak[1,3,4]

[1] Department of Human Genetics, University of California, Los Angeles, Los Angeles, California, USA

[2] Department of Cellular and Molecular Pharmacology, California Institute for Quantitative Biomedical Research, Center for RNA Systems Biology, and Howard Hughes Medical Institute, University of California, San Francisco, San Francisco, California, USA

[3] Department of Biological Chemistry, University of California, Los Angeles, Los Angeles, California, USA

[4] Howard Hughes Medical Institute, University of California, Los Angeles, Los Angeles, California, USA

Contact information:

Leonid Kruglyak
Gonda Research Center
695 Charles E. Young Drive South
Los Angeles, CA 90095
USA
Phone: 310-825-5486
Fax: 310-795-5446
Email: LKruglyak@mednet.ucla.edu

# Abstract

Heritable differences in gene expression between individuals are an important source of phenotypic variation. The question of how closely the effects of genetic variation on protein levels mirror those on mRNA levels remains open. Here, we addressed this question by using ribosome profiling to examine how genetic differences between two strains of the yeast *S. cerevisiae* affect translation. Strain differences in translation were observed for hundreds of genes. Allele specific measurements in the diploid hybrid between the two strains revealed roughly half as many *cis*-acting effects on translation as were observed for mRNA levels. In both the parents and the hybrid, most effects on translation were of small magnitude, such that the direction of an mRNA difference was typically reflected in a concordant footprint difference. The relative importance of *cis* and *trans* acting variation on footprint levels was similar to that for mRNA levels. There was a tendency for translation to cause larger footprint differences than expected given the respective mRNA differences. This is in contrast to translational differences between yeast species that have been reported to more often oppose than reinforce mRNA differences. Finally, we catalogued instances of premature translation termination in the two yeast strains and also found several instances where erroneous reference gene annotations lead to apparent nonsense mutations that in fact reside outside of the translated gene body. Overall, genetic influences on translation subtly modulate gene expression differences, and translation does not create strong discrepancies between genetic influences on mRNA and protein levels.

## Author summary

Individuals in a species differ from each other in many ways. For many traits, a fraction of this variation is genetic – it is caused by DNA sequence variants in the genome of each individual. Some of these variants influence traits by altering how much certain genes are expressed, i.e. how many mRNA and protein molecules are made in different individuals. Surprisingly, earlier work has found that the effects of genetic variants on mRNA and protein levels for the same genes appear to be very different. Many variants appeared to influence only mRNA (but not protein) levels, and vice versa. In this paper, we studied this question by using a technique called "ribosome profiling" to measure translation (the cellular process of reading mRNA molecules and synthesizing protein molecules) in two yeast strains. We found that the genetic differences between these two strains influence translation for hundreds of genes. Because most of these effects were small in magnitude, they explain at most a small fraction of the discrepancies between the effects of genetic variants on mRNA and protein levels.

# Introduction

Many genetic differences among individuals influence gene expression levels. Such regulatory variants are responsible for a large fraction of the variation in disease risk among humans and are also thought to be important for the evolution of phenotypes [1-3]. Regulatory variants can be mapped as expression quantitative trait loci (eQTL). Due to the relative ease and low cost of mRNA quantification, most eQTL studies have used levels of mRNA, rather than protein, as a measure of gene expression. The few studies that have used mass-spectrometry to examine genetic influences on protein levels reported surprisingly different genetic architectures for protein and mRNA levels [4-6]. For a given gene, many eQTL did not correspond to a protein QTL ("pQTL" [7]) and vice versa. Some analyses even suggested that eQTL and pQTL for certain groups of genes have significantly less overlap than expected by chance [5]. While more recent work [8-10] has found that eQTL and pQTL are more concordant than seen in the initial studies, numerous discrepancies remain. Together, these results have been taken to suggest that there must be substantial genetic variation acting on posttranscriptional processes.

Translation is an important determinant of cellular protein abundance ([11], but see [12]) and the rate of translation was shown to be a better predictor of protein levels than mRNA abundance [13]. Therefore, genetic variants that specifically influence translation are a potential explanation for the reported discrepancies between eQTL and pQTL.

Differences in gene expression between individuals can be caused by genetic variants that act in *cis* or by variants that act in *trans* [2]. Variants that act in *cis* influence the expression of alleles to which they are physically linked. In a diploid organism, *cis* acting variants can be detected as preferential expression of one allele compared to the other ("allele-specific expression", ASE) [14-19]. By contrast, *trans* acting variants influence the expression of both alleles of a gene to a similar extent.

Both *cis* and *trans* acting variants might have effects on translation. To affect translation in *cis*, a variant needs to reside within the mRNA of the given gene. By contrast, genetic variation in the various translation factors [20] might influence translation in *trans*. Further, mutations in ribosomal proteins can lead to highly specific differences in translation of small groups of mRNAs during mouse development [21], suggesting that genetic differences in genes beyond classic translation factors could affect translation in *trans*.

In this paper, we explored the influence of genetic variation on translation. We measured genome-wide translational activity in two genetically different strains of the yeast *S. cerevisiae* – the laboratory strain BY and the wine strain RM – as well as their diploid hybrid. Translation was measured by massively parallel sequencing of "ribosome footprints", i.e. of mRNA fragments that are associated with translating ribosomes [13,22]. By comparing the footprint data to measures of mRNA abundance gathered in parallel, we determined translation-specific influences on gene expression. In what follows, we distinguish three quantities. "mRNA abundance" quantifies RNA fragments from the polyadenylated transcriptome, irrespective of whether these molecules are translated. We denote as "footprint abundance" the number of RNA fragments bound by ribosomes, which is a measure of the total protein production for the given gene [13]. Finally, we refer to the ratio of footprint abundance to mRNA abundance as "translational efficiency" (TE) [13]). TE measures the extent to which the mRNA molecules of a given gene are translated.

We found that the differences in footprint abundance between BY and RM were highly correlated with the differences in mRNA abundance, both when comparing the parents and for ASE in the hybrid. Against this largely concordant backdrop, there were a small number of genes with evidence for strong translation-specific genetic effects on their expression, and hundreds of genes with more modest effects.

# RESULTS

*Differences in translation between two yeast strains*

We used ribosome profiling and mRNA sequencing to compare genome-wide patterns of translation in protein coding regions between the BY and the RM yeast strains. Alignment statistics are presented in Supplementary Tables S1 & S2 and discussed in Supplementary Note 1. There was excellent agreement between our measures in BY and those obtained by re-aligning the reads from a published yeast ribosome profiling dataset [13] (Supplementary Figure S1). This agreement is in spite of different growth media, slightly different strain backgrounds, several minor differences between library protocols, and substantially deeper sequence coverage in the current dataset (Methods).

mRNA abundance across the 6,697 genes annotated as coding or potentially coding in the yeast genome database varied by 3 – 4 orders of magnitude for the central 95% of genes. Footprint abundance spanned 4 – 5 orders of magnitude and, as expected, was highly correlated with mRNA abundance (Figure 1). TE varied by ~100 fold among the central 95% of genes, in line with previous observations in yeast [13]. Across genes, footprint abundance increased more rapidly than mRNA abundance such that genes with high mRNA abundance tended to have higher TE, while genes with lower mRNA abundance tended to have lower TE (Figure 1). This pattern is reminiscent of coordinated changes in mRNA levels and translation rates of yeast genes in response to diverse environmental stressors [23-26] as well as at steady state [27]. At the extreme end of this distribution were open reading frames (ORFs) categorized as "dubious" in the yeast genome database (Figure 1). Low or absent translation for dubious ORFs is consistent with the definition of ORFs in this category as "unlikely to encode an expressed protein" (www.yeastgenome.org). Together, these observations suggest that our data are of high quality and recapitulate known aspects of gene expression and translation.

mRNA and footprint levels in BY were highly correlated with those in RM (Figure 2A – B). Consequently, while 54% and 58% of genes had significant (binomial test, Bonferroni corrected $p < 0.05$) mRNA and footprint differences between the strains, more than 90% of these differences had small magnitudes of less than 2-fold (Table 1 &

Figure 2A – B). While we chose to employ the stringent Bonferroni multiple testing correction for the analyses presented in the main text, similar patterns were seen when using a more permissive threshold based on the false-discovery rate (FDR [28]) (Supplementary Table S3). That the majority of genes shows significant expression differences between BY and RM is in line with published estimates for mRNA levels – for example, 69% of genes were differentially expressed in a microarray based experiment [29].

To explore differences in translation between strains, we compared the mRNA differences to the footprint differences. The magnitudes of these differences were correlated (Spearman's rho = 0.71, Figure 3A), and this correlation became stronger when restricting the analyses to genes with a significant mRNA or footprint difference (rho = 0.77, Figure 3B). As expected from this correlation, genes with a significant mRNA difference were highly likely to also have a significant footprint difference (Fisher's exact test (FET): odds ratio = 4.6, $p < 2.2e-16$) and the direction of differences agreed for 82% (3,108 out of 3,798) of genes with a significant mRNA and / or footprint difference. Thus, a gene that differs significantly in mRNA abundance between BY and RM typically also has a significant footprint difference in the same direction. For most genes, translation carries forward mRNA differences to differences in protein synthesis. While a more conservative [30] significance testing method (the DESeq package, [31]) yielded substantially fewer differentially expressed genes (Table 1), the high agreement between mRNA and footprint differences remained intact (Supplementary Note S2 and Supplementary Figure S2A).

These observations leave open the possibility that translation may exert subtle, quantitative effects. Indeed, 42% of genes showed significant (Bonferroni corrected G-test $p < 0.05$) differences in TE, i.e., at these genes the ratio of footprint levels between strains differed from the respective mRNA ratio. As expected given the concordance between mRNA and footprint differences noted above, most of the TE differences were of small magnitude (Figure 2C & Table 1). Thus, while translation typically does not override mRNA differences between BY and RM, for many genes it subtly alters the degree to which mRNA differences are reflected at the level of protein synthesis.

*Allele specific translation in the BY / RM hybrid*

To investigate the extent to which translation is influenced by *cis*-acting variants, we gathered ribosomal footprint and mRNA data from the diploid hybrid between BY and RM and compared the expression of the two alleles. Reproducibility of the allele-specific measurements was high, as judged by comparing two biological replicates processed at the same time (Supplementary Figure S3). As expected, the number of genes with significant ASE was less than the number of genes with differences between the BY and RM parental strains (Figure 2 & Table 1): we detected significant (Bonferroni corrected binomial test $p < 0.05$) ASE for the mRNA levels of 6% of genes and for the footprint levels of 6% of genes. The mRNA estimate is lower than previous estimates for these two yeast isolates (~14% [17] – ~20% [32]) because of the stringent Bonferroni correction we applied here; when we use a FDR-based cutoff we obtain higher numbers of genes with significant ASE (Supplementary Table S3).

Variants that act in *cis* to alter translation should result in footprint ASE that is not a direct reflection of mRNA ASE. By contrast, we found that the magnitudes of mRNA and footprint ASE were correlated (rho = 0.72) when considering genes with significant mRNA or footprint ASE (Figure 3E & Supplementary Figure S2B). Consequently, genes with significant mRNA ASE were very likely to also have significant footprint ASE (FET odds ratio = 25, $p < 2.2e-16$) and 83% (260 / 312) of genes with significant mRNA or footprint ASE agreed in the direction of ASE. Thus, allele specific footprint levels mostly reflect allele specific mRNA expression, suggesting that *cis* acting variants with strong effects on translation are rare in BY and RM.

To search for genes that may carry *cis* acting variants that influence translation, we tested for allele-specific TE, i.e. for genes where the ratio of mRNA ASE differs from the ratio of footprint ASE. Significant (G-test, Bonferroni corrected $p < 0.05$) allele-specific TE was found for 3% (n = 106) of genes (Table 1). While most of these effects had small magnitude (Figure 2F), 26 genes had significant allele-specific TE greater than 2-fold (Table 2). For three of these genes, allele-specific TE is due to premature translation termination in one strain compared to the other (Table 2; see Supplementary Table S4 and Supplementary Figure S2C for genes with allele-specific TE differences identified using DESeq). The remaining 23 of these 26 genes appear to carry *cis* acting variants that substantially alter the translation rate without disrupting gene structure.

Cis *and* trans *contributions to translation*

The overall contribution of *cis* vs. *trans* acting variants to gene expression variation can be determined by comparing differences in expression between two strains to allele-specific expression in their diploid hybrid [14,17]. For genes that are entirely regulated in *cis*, the parental difference should be completely recapitulated by allele-biased expression in the hybrid. For genes that are entirely regulated in *trans*, expression of the two alleles should be the same in the hybrid, irrespective of the parent difference. Consequently, in the absence of noise, if all genes in the genome were exclusively affected by *cis*-acting variants, the slope of the relationship between allelic differences and parental differences should equal one, and if all genes were exclusively affected by *trans*-acting variants, the slope should equal zero.

In our data, the slopes of these relationships (calculated using major axis estimation [33]) were 0.35 for mRNA and 0.37 for footprint abundance (Figure 4A). Because the data are noisy, we used bootstrap analysis to show that these estimates are significantly different from zero or one (Figure 4C), implying the presence of both *cis*- and *trans*-acting variants. In order to avoid potential biases associated with different quantifications (all mapped reads in parents vs. only reads overlapping SNPs in the hybrid), the parental fold changes used in these analyses were based on allele counts at the same set of SNPs used to analyze the hybrid. Bootstrapped distributions of the mRNA and footprint slope estimates overlapped substantially (Figure 4C). Thus, the relative contributions of *cis*- and *trans*-acting variants on mRNA abundance are faithfully represented in footprint abundance.

Next, we asked if the effects of previously identified eQTL (i.e., individual loci identified through their effects on mRNA levels) are reflected in our data. We stratified genes according to whether they were influenced by a local (likely *cis*-acting) eQTL, a distant (likely *trans*-acting) eQTL, or by both types of eQTL, using eQTL reported in [29]. Genes with local eQTL had a significantly steeper slope (the 95% confidence intervals from 1,000 bootstraps did not overlap) than genes with distant eQTL (Figure 4B & C). Thus, the effects of known genetic variants are recapitulated in our mRNA data. The effects of known eQTL were also seen in footprint abundance (Figure 4B & C). Together, these analyses suggest that the relative importance of *cis* vs. *trans* acting genetic variation on footprint abundance is largely similar to that on mRNA abundance.

*Relationship between mRNA differences and translation differences within and between yeast species*

There are several possible relationships between mRNA differences and footprint differences (Figure 5A). First and most obviously, mRNA and footprint differences can be statistically indistinguishable; these genes fall near the diagonal in Figure 5A and will not be considered further. When TE differs significantly between BY and RM, this can have the following effects. A significant mRNA difference can correspond to a larger (reinforced) or smaller (buffered) significant footprint difference in the same direction, no significant footprint difference (complete buffering), or a footprint difference in the opposite direction ("inverted" genes). Finally, a significant footprint difference can appear in the absence of a significant mRNA difference ("FP only" genes).

To examine the relative frequencies of these scenarios, we partitioned the genes with significant TE differences between BY and RM ("TE genes", 42% of genes using the Bonferroni significance cutoff) (Table 3, Figure 3C & Figure 5B & C). For 5% of the TE genes, neither the mRNA nor the footprint difference was significant, providing little further information. Thirty-one percent of the TE genes showed reinforcement and 27% of the TE genes had a footprint difference in the absence of an mRNA difference. Conversely, 29% of the TE genes showed partial or complete buffering. Only 7% of the TE genes fell in the "inverted" category. The results for the hybrid data are given in Table 3 and Figure 5B & C. Results based on a less stringent FDR significance cutoff as well as those based on the more conservative DESeq significance testing framework [31] are shown in Supplementary Table S5.

Recent work comparing the two yeast species *S. cerevisiae* (two isolates of which we study here) and *Saccharomyces paradoxus* reported an excess of directional effects of translation. An mRNA difference between species was typically accompanied by a difference in translation in the opposite direction, resulting in a smaller or inverted footprint difference [34,35]. We sought to test if this excess also exists between BY and RM. Because the results of this inference could depend on the precise choice of how mRNA differences and footprint differences are compared, we systematically tabulated the four possible comparisons of directional effects of translation (Figure 5 and Tables 3 & 4). These four comparisons arise by including or excluding the "FP only" genes, and by including or excluding the "inverted" genes. The detailed results are presented in Supplementary Note S3. Taken together, they suggest that variation in translation between BY and RM more often increases than decreases footprint differences compared to mRNA differences, although this conclusion depends on the precise way in which TE genes are compared.

To ensure a consistent comparison with these results between BY and RM, we reanalyzed the inter-species results provided in [34] as well as p-values and fold-changes kindly provided by the authors of [35] (Table 4, Figure 5B & C, Supplementary Table S6). In line with the published results for the McManus *et al.* dataset [34], there was a strong excess of opposing over increasing effects of translation in both the parent and the hybrid data. This effect was robust to the precise choice of tested gene groups (Table 4). In the hybrid data from Artieri & Fraser [35], we replicated the reported excess of opposing genes when compared against reinforced genes. This result was less robust to the precise choice of comparison (Supplementary Note S3). Overall, the signal of translational buffering is strong in one of the two published datasets and more dependent on the precise analyses in the other dataset. Discussion of additional analyses in the interspecies comparisons is provided in Supplementary Note S3 and Supplementary Figure S4.

*mRNA and footprint levels are correlated with genetic influences on protein levels*

Next, we asked to what extent the mRNA and footprint differences between BY and RM correspond to differences in protein abundance that are due to individual genetic loci. We recently used a bulk-segregant approach to map *trans*–acting pQTL in BY and RM [8]. The high statistical power of that approach resulted in the identification of multiple pQTL for many of the analyzed genes. Of the 160 genes in our earlier study, 114 could be analyzed in the present parent and hybrid datasets. For each of these 114 genes, we summed the allele frequency differences at the pQTL, a measure related to the genetic effects of the pQTL (Methods). These summed measures provide a rough expectation of protein level differences between BY and RM that are due to *trans*-acting variation.

As expected given that all the pQTL considered here act in *trans*, there was no correlation between the predicted protein differences and ASE in the hybrid data (Figure 6C & D). By contrast, we found that the summed pQTL effects correlated significantly with differences in mRNA abundance between BY and RM (Spearman rank correlation rho = 0.36, p = 0.0001; Figure 6A). Thus, the aggregate effects of loci that were detected through their effects on protein levels are reflected in mRNA differences between strains. There was a slightly stronger correlation between the summed pQTL effects and the strain differences in footprint abundance (rho = 0.46, p = 3e-7; Figure 6B), but this difference in the strength of correlations appeared to be primarily driven by a few genes (112 of 1,000 bootstrapped datasets showed a larger mRNA than footprint correlation). These observations further support the hypothesis that most pQTL arise from genetic influences on mRNA levels, perhaps augmented by minor additional contributions by genetic effects on translation.

*Effects of nonsense and frameshift mutations on translation*

The single base pair resolution of ribosome-profiling data permits detailed examination of footprint abundance along the length of a gene, and of how these patterns are affected by genetic variation. In particular, variants that create or disrupt a stop codon are of interest due to the potentially large phenotypic consequences. In addition, insertions and deletions (indels) that lead to a shift in the reading frame of a coding gene will usually result in premature translation termination or, more rarely, in translation proceeding beyond the original stop codon. All these types of variants should lead to detectable differences in the pattern of footprint coverage.

Among our set of high quality coding SNPs in 3,376 genes that had a status of "verified" according to the SGD database, we identified 18 sites where RM relative to BY has gained a premature stop codon and 10 sites where RM has lost the annotated stop codon (Supplementary Data S1). In addition, we catalogued 32 short indels predicted to lead to a frameshift. We visually examined the footprints patterns in BY and RM at these sites. Data tracks showing these patterns across the genome will be made available for interactive browsing in the UCSC browser.

Of the 18 gained stops in RM, three were in genes that were not expressed and could not be analyzed. Of the 15 sites in expressed genes, nine led to clearly visible premature termination (Figure 7A for an example). Four putative premature stop codons were located in a part of the ORF that, while annotated as part of the coding sequence, is in fact upstream of the region we found to be translated in both BY and RM (Figure 7B). These four sites therefore do not affect the protein, but reflect errors in gene annotation. The remaining two sites were situated in the translated part of the coding sequence, but did not lead to a visible reduction in translation. Closer inspection of these two SNPs showed that both of them are part of multi-base substitutions that together lead to an amino acid substitution instead of a nonsense mutation. Thus, only 60% (9 / 15) of our list of predicted nonsense mutations in expressed genes had detectable effects on protein sequence. We note that six of these truncating mutations were close to the 3' end of the coding sequence, where they may be less likely to severely disrupt protein function [36].

Of the ten sites where a BY stop codon was absent in RM, three resided in genes with no or very low expression. Four did lead to visible ribosomal readthrough, and two of these sites are in fact known instances of difference in primary protein structure between BY and RM. For example, the gene *NIT1* and the gene annotated to lie immediately downstream (YIL165C) form a single ORF in other yeast species [37] and other *S. cerevisae* strains [38]. The two genes are annotated as two separate ORFs because the yeast genome annotation is primarily based on BY, which carries a premature stop codon inside the ORF (Figure 7C). The remaining three lost stop codons did not visibly result in translational readthrough. Closer examination of the sequence context reveals that these codons are immediately followed by a secondary stop codon that compensates for the lost stop codon [39].

We made similar observations for the 32 putative frame shifting indels: ten were in genes with low expression, five were in untranslated regions erroneously annotated as coding, three were in repetitive regions and may be due to alignment errors, six were close to the end of the ORF, one was downstream of a premature stop and therefore of no consequence itself, and only seven led to visible early termination or extension of the frame-shifted protein.

# DISCUSSION

We used ribosome profiling [13] to explore how genetic differences between the two yeast strains BY and RM influence mRNA abundance and translation. We found that most genes with significant differences in mRNA levels had footprint differences in the same direction. Thus, translation typically carries forward genetic influences on mRNA levels into differences in protein synthesis. While we did detect hundreds of genes that showed evidence for genetic effects on translation, most of these effects subtly modulate rather than override mRNA differences. Genetic variants that induce strong, specific effects on translation appear to be infrequent in BY and RM.

We made similar observations in the hybrid between BY and RM. Significant allele-specific mRNA expression was highly correlated with allele-specific footprint abundance. Therefore, with a few exceptions (e.g. those listed in Table 2 and Supplementary Table S4), most genes do not carry *cis*-acting variants that have large, specific influences on translation.

By comparing the parental differences to ASE in the hybrid [14], we found that the relative contribution of *cis*- vs. *trans*-acting variants on footprint levels was similar to that on mRNA levels. Further, individual local and distant eQTL that had earlier been identified based on their effects on mRNA levels [29] influence the *cis* vs. *trans* contribution in both the mRNA and footprint data presented here. These eQTL therefore are carried forward to translation and would be expected to also affect protein levels.

Analyses of a mass spectrometry dataset have reported substantial discrepancies between genetic influences on mRNA and protein differences between BY and RM [4,5]. Our ribosome profiling data provides little evidence that genetic effects on translation might be responsible for these discrepancies. This observation is in line with recent pQTL studies in yeast that leveraged improvements in protein measurements and experimental design [8,9,40] and found that eQTL and pQTL are not as discordant as reported previously. To the extent that the remaining discrepancies between eQTL and pQTL are real (as opposed to, for example, due to experimental variation [41]), our results here suggest that they are more likely caused by genetic influences on protein degradation rather than on translation.

Two recent papers examined the evolution of mRNA and footprint levels between the yeast species *S. cerevisiae* and *S. paradoxus* [34,35]. Both studies reported that mRNA differences are more often opposed than reinforced by translation. Motivated by these reports, we conducted similar analyses in our data. We found some evidence that genes with strain differences in TE between BY and RM tend to more often have footprint differences larger than the corresponding mRNA differences, the opposite pattern of what was reported for the species comparisons. A similar pattern was recently observed for allele-specific translation in *Candida albicans* [42]. However, in BY / RM, this pattern was dependent on the precise fashion in which the analysis is performed (Table 4 & Supplementary Note S3), in line with the observation that the mRNA and FP differences are similar in magnitude (Figure 3). Given that this inference in BY and RM was dependent on the exact way in which TE genes are grouped, we performed the same set of comparisons in the two published interspecies data sets. In one of the two studies, the reported excess of opposing effects of translation was robust across comparisons, while in the other study the results were more ambiguous.

In our opinion, none of the groupings of the TE genes we used to compare the directional effects of translation is obviously more correct than the others. For example, while it would be our preference to include genes with a footprint but no mRNA difference in the analyses, these genes were excluded in the published inter-species analyses that only analyzed genes with a significant mRNA difference [34,35]. It is also unclear whether (and where) genes with a footprint difference that is inverted compared to the mRNA difference should be included. In addition, differences in the precise experimental protocols between studies may contribute to the different results. For example, the technical variance in footprints is typically higher than that in mRNA, and also differs among the different datasets (Supplementary Figure S5). We are thus hesitant to draw strong conclusions about the relative importance of opposing / buffering or reinforcing / increasing effects of translation within and between yeast species, although we cannot rule out genuine evolutionary differences between these intra- and interspecific comparisons.

Which cellular mechanisms might explain the observed cases where translational differences reinforce, buffer or invert an mRNA difference? The simplest explanation is that these cases involve two or more variants: one altering mRNA levels and another altering translation. Alternatively, more parsimonious explanations might involve a single mutation that affects both mRNA levels and translation rates. There is growing evidence for coordination among the stages of gene expression [43]. For example, positive correlations between mRNA abundance and translation rates have been observed during unperturbed growth [27], as well as for mRNA abundance changes in response to various stressors [23-26]. Recent evidence suggests that in addition to transcription, promoter sequences influence the subcellular localization and translation rates of yeast mRNAs [44]. While the precise mechanisms that mediate these coordinated effects are not fully understood, there is some evidence that the transcription machinery can influence the translational fate of mRNAs through the RNA polymerase II subunits Rpb4 and Rpb7 [45].

Translation can also stabilize mRNA molecules by protecting them from degradation [27,46], so that a higher translation rate *per se* can result in higher mRNA levels at steady state. A sequence variant that increases TE of a given gene could then not only result in higher footprint levels but also increase mRNA levels, even if the variant has no effect on transcription. Careful study of the dynamics of translation (e.g. [47]) will be needed to further address this question.

Our analyses of nonsense and frameshift polymorphisms showed that these variants indeed result in detectable differences in translation. However, the results serve as a reminder to exercise caution when interpreting the potential functional impact of variants identified in next generation sequencing datasets, especially for variants with putative large effects [36]. Sequence context (e.g. secondary stop codons downstream of a lost stop [39]) and multi-base substitutions can obscure the true consequences of a variant called from a high-throughput pipeline when considered in isolation. Further, even in an extremely well annotated genome such as that of *S. cerevisiae*, errors in gene annotation can generate the illusion of severe differences in protein sequence between strains when in fact the corresponding variants reside outside of the coding region. Our list of variants between BY and RM with validated effects on translation as well as of problematic gene annotations (Supplementary Data S1) can be useful to assess the consequences of genetic differences between these yeast strains. We have also made available tracks in the UCSC genome browser [48] that allow easy visualization of translation patterns between BY and RM for any gene.

Molecular phenotypes such as mRNA and protein levels (as well as others [49,50]) provide crucial intermediates for connecting DNA sequence variation to organismal phenotypes. New measurement technologies will allow an increasingly fine-grained view of the mechanistic connections between the levels of molecular traits and illuminate how genetic variation shapes organisms.

# MATERIALS AND METHODS

*Yeast strains*

We studied the same strains as in Bloom *et al.* [51]. The common laboratory BY strain we used had mating type *MAT**a***. The RM strain was originally isolated from a vineyard. Our RM strain had genotype *MATα hoΔ::hphMX4 flo8Δ::natMX4 AMN1-BY*. Both strains were prototrophic, *i.e.* they did not carry any engineered deletions of metabolic genes. These deletions are commonly used as genetic markers that can have strong effects on gene expression [1]. The haploid parental strains were crossed to generate the diploid hybrid. BY and RM differ in cycloheximide resistance at a dose several orders of magnitude lower than those used in the ribosome profiling protocol [13,51]. To confirm that the parents and the hybrid were equally sensitive to the high cycloheximide dose used here to block translation, we attempted to grow them at 30°C in triplicates in liquid yeast nitrogen base (YNB) medium with a range of cycloheximide concentrations centered on the dose used in the ribosome profiling protocol. While growth was normal in negative controls without cycloheximide, there was no growth within 24 hours in any of the cycloheximide doses tested.

*Ribosome profiling and sequencing*

Libraries for RNA-seq and ribosome profiling were prepared as described in [13], with the following exceptions: (1) cells were cultured in YNB, (2) the reverse-transcription step was primed by ligating miRNA Cloning Linker 1 (IDT) onto the RNA fragments, and (3) highly abundant rRNA species were hybridized to biotinylated oligos and subtracted using streptavidin-coated DynaBeads (Invitrogen) as in [52]. Deep sequencing was performed on the Illumina HiSeq 2000 platform. Raw reads are available in the NCBI Gene Expression Omnibus under accession GSE55400.

*SNP set for allele specific quantification*

We employed a set of filters to ensure unbiased estimates of ASE. We used the program BWA [53] to align high coverage (> 50X) 94 bp paired-end whole genome Illumina sequencing data from the BY and the RM strains used in this study [51] to the reference yeast genome version sacCer3 downloaded from the UCSC genome browser ( http://genome.ucsc.edu ). We used a custom python script kindly provided by Martin Kircher to remove PCR duplicates. Samtools [54] was used to extract a preliminary set of SNPs with variant quality score > 30 and with an estimated alternative allele frequency of 1 ("AF1=1" flag in the vcf file). Next, we retained only biallelic SNPs where our RM strain carries an alternative allele and our BY strain carries the genome reference allele. There were 43,154 SNPs in this initial set.

We sought to restrict this set to those SNPs where short sequencing reads (such as those obtained in ribosome profiling) can be aligned to unique positions in both the BY and the RM reference genome. For each SNP, we extracted the 30 bp up- and downstream sequence from the BY genome reference (sacCer3), from both the plus and the minus strand. The SNP allele itself was set to the RM allele. The resulting 61 bp sequences were aligned to the RM reference genome downloaded from the Broad Institute (http://www.broadinstitute.org/annotation/genome/saccharomyces_cerevisiae.3/Info.html ) using BWA [53]. We removed SNPs whose flanking sequences mapped to more than one position in the RM genome as well as SNPs where multiple SNPs mapped to the same position in the RM genome. The number of SNPs after these filters was 38,706.

Next, we sought to remove SNPs with alignment biases towards one or the other reference genome by examining the alignment behavior of publicly available DNA sequence data obtained from a BY / RM hybrid [55]. Any allelic bias seen in hybrid DNA sequences necessarily is of technical origin and indicates problematic SNPs. We trimmed the hybrid DNA reads to 30 bp single end, aligned them to both the BY and RM reference genomes and counted the number of reads that overlapped the BY or RM reference alleles at each SNP, exactly as described below for our mRNA and footprint reads. To identify SNPs with allelic bias beyond that expected by chance, we simulated an unbiased dataset as follows. For each SNP, we generated allele counts assuming a binomial distribution with p = 0.5, and at a depth of coverage drawn from the observed data. We determined criteria for SNP exclusion based on visual comparison of the observed hybrid DNA dataset to the simulated unbiased data. We removed SNPs with very high (> 100 fold) and very low (< 30 fold) coverage, as well as any remaining SNPs where the frequency of the BY (and, equivalently, the RM) allele was less than 0.3 or more than 0.7. After these filtering steps, 36,089 SNPs remained.

We noted a population of SNPs with hybrid DNA allelic ratio centered at ~1/3, i.e. a 2:1 bias towards the RM genome. Further inspection revealed that these SNPs all resided in regions where DNA sequencing coverage in our RM parent was twice as high as that in our BY parent. Nearly all of these regions were situated at chromosome ends and likely reflect segmental duplications of these distal regions in the RM strain compared to the BY reference genome. These regions extended for several kb and contained annotated protein coding genes, in line with the recent observation that subtelomeric regions contain large structural variants that segregate among wild yeast [56]. We visually examined the coverage across our BY and RM parent DNA sequences and excluded any regions with evidence for segmental duplications in the RM but not the BY parent. This removed 821 SNPs, for a remaining set of 35,268.

Finally, because we are interested in quantifying expression of protein coding genes, we retained only the 23,412 SNPs in ORFs annotated in the SGD database (www.yeastgenome.org, accessed on 06 / 28 / 2013). SNPs in ORFs annotated as overlapping on the same strand were removed. However, because the mRNA and footprint data are strand-specific, we were able to retain 395 SNPs that overlap ORFs on different strands for a total of 23,807 quantifiable positions in 4,462 ORFs (Supplementary Data S2).

*Read processing and alignments*

Because the reference yeast genome is based on a strain with the BY background, sequence differences between the reference and RM make read alignments from an RM sample more difficult, especially with short reads such as the ~32 base pair (bp) ribosomal footprint fragments. To counter this problem, we implemented a computational pipeline that uses "personalized" genome references for the BY and the RM strain to allow unbiased read mapping.

Prior to mapping, we removed sequences corresponding to the Illumina adapter sequence (CTGTAGGCACCATCAAT) opposite the sequencing priming site and discarded all reads that did not contain these adapter sequences. We also removed the first base from each read as these often corresponded to adenosines introduced during ligation in the library preparation protocol.

The trimmed reads were mapped using BWA [53] as follows (see Supplementary Table 1 for alignment statistics). For the comparisons between the BY and RM strain, reads can be considered irrespective of whether they cover a SNP or not. Reads from the BY strain were mapped to the BY reference genome (version sacCer3). Reads from the RM strain were mapped to a modified version of the BY reference where the 43,154 SNPs between BY and RM as described the section above were set to the RM allele. The rationale for using this strategy was to maximize the number of RM reads that can be mapped to the BY reference without penalizing reads that contain a sequence difference between BY and RM, while still being able to directly use the BY gene annotations. We counted only uniquely mapping reads on the correct strand in genes.

For the ASE analyses, we are only interested in reads that span a SNP between the BY and RM strains. We noted that the short reads produced in ribosome profiling are heavily biased against mapping RM reads to the BY reference (not shown). We therefore mapped all reads to both the BY reference and the RM reference available from the Broad Institute. We considered only reads that mapped to one of these two reference sequences uniquely and without mismatch. This strategy guarantees that reads that span a sequence difference between BY and RM can be unambiguously assigned to the parental chromosome they originated from. At each of the 23,807 high quality coding SNPs (s. section above), we counted the number of reads that mapped to the correct strand of the BY or the RM genome. When a read overlapped multiple closely linked SNPs, it was randomly counted towards one of them. Because we excluded reads with mismatches, our strategy excludes all reads with sequencing errors. For comparison of ASE in the hybrid to differences between the parent strains, we re-mapped the BY and RM parent reads and quantified allele-specific expression as described in this section.

*Quantification of mRNA and footprint abundance*

For determining the genomic source of reads in the libraries (ORFs, UTRs, ncRNAs, etc.) as well as for the comparison between the parent strains reported in the main text, we used htseq-count (http://www-huber.embl.de/users/anders/HTSeq/doc/overview.html) and annotations extracted from SGD (www.yeastgenome.org).

For the analyses of allele specific expression, we added the allele counts for all SNPs in a gene. For the hybrid, we summed the counts from the two replicates, with the exception of reproducibility analyses. While all statistical analyses were performed directly on count data (s. below), the figures show gene abundance as log10-transformed fractions of total counts for the given sample. Translation efficiency (TE) for a gene was calculated as the difference between the log10-transformed mRNA fraction and the log10-transformed footprint fraction. All quantifications, both for whole ORF and SNPs, are available in Supplementary Data S2.

*Statistical analyses*

All statistical analyses were performed in the R programming language (www.r-project.org). Unless stated otherwise, we calculated slopes using major axis estimation [17,33] as implemented in [57]. Correlations were calculated as nonparametric Spearman's rank correlations to avoid making assumptions about the distributions of the data.

We used two different count-based approaches to gauge statistical significance. In the main text, we report the results from binomial tests while in Supplementary Note S2 we describe results obtained with the DESeq analysis framework [31].

Count-based binomial tests have higher power when the absolute number of counts is high. The number of reads mapped was different between different samples and between different data types. Specifically, the parental footprint libraries had 30% - 70% more reads than the parental mRNA libraries (Supplementary Table S1). We removed these differences in total read counts by downsampling as follows, largely following the procedures described in [58]. For each comparison (parental analyses based on all mapped reads or allele-specific analyses counting only reads spanning SNPs), we identified the sample with the lowest total number of counts. This sample remained as observed, while in the other samples we randomly sampled read counts to match the smallest library size. We call this dataset the "downsampled" dataset and the unadjusted data prior to downsampling the "raw" dataset.

To test for differential expression of mRNA or footprints, we generated a second datatset from the downsampled data using hypergeometric resampling [58]. The goal is to avoid differences in power to call differential mRNA or footprint expression when the absolute counts are different for these two data types. For example, if a gene is highly transcribed but has a low rate of translation, the mRNA data might have more counts and therefore higher power. For each gene, we grouped the samples into pairs of samples to be compared. These pairs are the counts from the two alleles in the hybrid, or the counts from the respective parental samples. mRNA and footprint counts formed separate pairs. For each gene, we identified the pair with the smallest sum of counts. This pair remained as observed. For the other pairs, we used hypergeometric sampling to generate data with the same sum of counts as the pair with the lowest sum of counts. We call the resulting dataset the "hypergeometric" dataset.

We removed all genes from the analyses where, in the hypergeometric dataset, any pair had a sum of counts < 20 (following [58]), and all genes where any individual sample had a count of zero (these two criteria are not redundant because in a few cases, one member of a pair had > 20 counts while the other had zero counts). Genes that did not satisfy this filter were also excluded from the downsampled dataset. After filtering, there were 5,316 and 3,342 genes available for analysis in the parent dataset and the SNP-based hybrid dataset, respectively. No genes were removed from the raw dataset.

For the analyses of reproducibility in the hybrid we performed the same down-sampling approach but kept the two hybrid replicates as separate samples.

We tested for differential mRNA of footprint expression using binomial tests on the "hypergeometric" dataset. For most analyses the p-values were adjusted for multiple tests using Bonferroni correction. We also estimated false discovery rates by calculating q-values [28,59]. For DESeq (s. below), multiple tests were corrected using the Benjamini-Hochberg procedure [60] that is the default in DESeq [31].

To test for differential TE, we used the "downsampled" data to test if the ratio of footprint and mRNA counts differed between strains. This test cannot be performed in the "hypergeometric" data because in these, mRNA and footprints have been sampled to the same level, overriding the TE signal. We used G-tests on 2x2 tables of the form:

|    | Footprints | mRNA |
| --- | --- | --- |
| BY | $y_{i,\ BY\ footprints}$ | $y_{i,\ BY\ mRNA}$ |
| RM | $y_{i,\ RM\ footprints}$ | $y_{i,\ RM\ mRNA}$ |

where $y_{i,strain}$ is the number of downsampled counts for gene $i$ in *strain* (BY or RM).

As an independent approach to test for statistical significance, we used the DESeq analysis framework [31]. DESeq models the counts using a negative binomial distribution and asks if, for a given gene, the observed mean difference between strains is more than expected given the variance for a gene of the given abundance. As such, DESeq takes into account the fact that more highly expressed genes have higher counts and therefore higher statistical power than less abundant genes.

We used DESeq version 1.16 on raw count data, i.e. without any prior normalization, but excluding genes where all samples had a count of zero. In the parent data and hybrid, 6,697 and 4,462 genes were available for analysis in DESeq. To analyze the BY and RM parent data, we calculated dispersion factors using the method="blind" option together with the sharingMode="fit-only" option to approximate experimental (noise) variance by treating the BY and RM samples as "replicates" as recommended in the DESeq manual. This procedure is known to be conservative [31].

In the hybrid data, we used the allele counts from the two replicates separately (without any normalization, and without summing the replicates) and compared them using the default DESeq settings. Further, in the hybrid data it is possible to use DESeq to test for differential TE by using the nbinomGLMTest() function to compare models specified by the formulae:

*H1: count = molecule + strain + molecule:strain*

and

*H0: count ~ molecule + strain*

Where *count* is the count data, *molecule* is either mRNA or footprint, *strain* is either BY or RM, and *molecule:strain* indicates an interaction term. In words, this test asks for each gene whether the ratio of mRNA to footprint counts is different for the two alleles in the hybrid.

All data and results, including raw, downsampled and hypergeometric count datasets, p-values and fold changes, are available in Supplementary Data S2.

*Comparison of mRNA and footprint differences to pQTL effects*

Recently, we showed that the expression of many proteins is influenced by multiple loci that segregate between the BY and the RM isolates [8]. The Albert *et al.* and the present datasets overlap for 114 proteins when considering only genes that can be analyzed in the hybrid data (i.e., that are expressed and that contain at least one SNP). To generate a rough expectation for the aggregate effect that the multiple pQTL have on a given protein, we added their effects. The pQTL in our earlier study were obtained by comparing allele frequencies in populations of cells with high and low protein expression, so that direct estimates of QTL effects (i.e. the expected magnitude by which protein expression differs between the different alleles) are not available. Instead, we used the observed difference in allele frequency at the pQTL location as a measure of effect size. Note that the locus effects can cancel each other: two pQTL with the same absolute allele frequency difference, but with opposite sign will result in an expected aggregate effect of zero. The summed pQTL effects were compared to mRNA and footprint differences from the present study using nonparametric Spearman rank correlation.

To test whether the footprint differences or the mRNA differences correlated better with the pQTL effects than the mRNA differences, we constructed 1,000 bootstrap datasets. In each of these datasets, we randomly sampled from the 114 genes with replacement and calculated both the mRNA and footprint correlations. We calculated the p-value as the fraction of bootstrap datasets where the mRNA difference / pQTL effect correlation exceeded the footprint difference / pQTL effect correlation.

*Published footprint data for* S. cerevisiae *and* S. paradoxus

For Artieri & Fraser [35], we used p-values and fold changes kindly provided by the authors. P-values had been calculated as reported in [35]. We excluded all genes where any of the mRNA, footprint or TE p-values could not be calculated due to low coverage or due to the two replicates not agreeing in direction of effect. This resulted in 1,861 genes available for analysis in the species comparison and 1,451 genes available in the hybrid comparison. We corrected the p-values for multiple testing by calculating q-values [28] and treated q-values of < 0.05 as statistically significant.

For McManus *et al.* [34] we used the read counts, p-values and fold-changes provided in their Supplementary Table S5. There were 4,863 genes available in both the parent and hybrid analyses of the McManus *et al.* data. The p-values reported in [34] are already corrected for multiple testing and we deemed p-values < 0.05 as statistically significant.

*Analyses of nonsense and frameshift mutations*

We restricted these analyses to coding genes with a "verified" status according to the SGD database. For nonsense SNPs, we considered only the set of high-quality SNPs we used in our ASE analyses. For indels, we considered indel calls produced by samtools mpileup [54]. Because our sequencing coverage of the BY and RM strains is very deep and often exceeds the default limit of 250 fold coverage, we used mpileup with parameters –d 1000 and –L 1000. The resulting indels were further filtered using the bcftools vcfutils.pl varFilter script to only retain indels with coverage of at least 10-fold, variant quality of at least 30, an estimated allele frequency in the sample of 1 (as the sequenced strain was haploid). Finally, we only considered indels in "verified" genes. To identify nonsense SNPs and frameshift indels, we used the standalone perl version of the Ensembl Variant Effect Predictor tool [61] with Ensembl cache data for the yeast sacCer3 genome build available in Ensembl build 72. To examine the effects on translation of the resulting nonsense and frameshift variants, we generated custom tracks in the bedGraph format for display in the UCSC genome browser [48]. These tracks show the start coordinate of each read. The tracks will be made available for interactive browsing in the UCSC browser. The track plots in Figure 7 were generated using the Gviz R package [62].

## Acknowledgements

We are grateful to Hunter Fraser, Carlo Artieri, Joel McManus, Joshua Bloom, Sebastian Treusch, J.J. Emerson, Casey Brown, three anonymous reviewers and members of the Kruglyak lab for discussions on the analyses and comments on the manuscript.

# Tables

Table 1 – Differential expression statistics

| Comparison | Reads | Data | Analyzed genes | 2-fold | Binomial / G test[1] | Intersect | DESeq[2] |
|---|---|---|---|---|---|---|---|
| Parent | All | mRNA | 5,316 | 331 | 2,862 | 314 (6%) | 189 |
| Parent | All | Footprint | 5,316 | 514 | 3,057 | 490 (9%) | 145 |
| Parent | All | TE | 5,316 | 135 | 2,228 | 111 (2%) | NA |
| Parent | SNP | mRNA | 3,342 | 249 | 517 | 171 (5%) | 75 |
| Parent | SNP | Footprint | 3,342 | 475 | 671 | 289 (9%) | 67 |
| Parent | SNP | TE | 3,342 | 329 | 319 | 97 (3%) | NA |
| Hybrid | SNP | mRNA | 3,342 | 100 | 198 | 40 (1%) | 40 |
| Hybrid | SNP | Footprint | 3,342 | 194 | 210 | 67 (2%) | 70 |
| Hybrid | SNP | TE | 3,342 | 216 | 106 | 26 (1%) | 9 |

[1]Bonferroni corrected $p < 0.05$. Binomial tests were run on mRNA and footprint data. G-tests were used to test for differential TE.

[2]DESeq was run on all of genes where at least one sample had more than zero counts (6,457 genes for the parent comparison using all reads and 4,361 genes for the SNP-based analyses). DESeq results were corrected for multiple testing using the Benjamini-Hochberg correction.

Table 2 – Strong *cis* effects on translation

| Gene | TE log2(fold change) | p-value | mRNA log2(fold change) | p-value | FP log2(fold change) | p-value | Notes |
|---|---|---|---|---|---|---|---|
| YAL054C (*ACS1*) | -3.81 | 1.2E-05 | 0.56 | 1 | -3.25 | 2.2E-04 | |
| YBR107C (*IML3*) | 1.68 | 2.8E-10 | -1.29 | 5.4E-06 | 0.39 | 0.1 | |
| YBR114W (*RAD16*) | 1.10 | 8.3E-06 | -0.86 | 0.01 | 0.25 | 0.2 | |
| YDL231C (*BRE4*) | -3.41 | 1.8E-10 | -0.23 | 0.02 | -3.63 | 2.5E-08 | |
| YDL203C (*ACK1*) | 1.00 | 1.3E-09 | -0.74 | 2.9E-06 | 0.27 | 6.1E-02 | |
| YDL124W | 1.01 | < 2e-16 | 1.16 | 2.1E-38 | 2.17 | 1.0E-107 | |
| YDR133C | -2.48 | < 2e-16 | -1.20 | 1.0E-152 | -3.68 | < 2e-16 | (1) |
| YEL066W (*HPA3*) | 1.53 | 4.3E-10 | -0.43 | 0.02 | 1.10 | 1.2E-09 | |
| YGL252C (*RTG2*) | -1.05 | 4.9E-06 | 0.23 | 0.5 | -0.82 | 1.0E-05 | |
| YGL163C (*RAD54*) | -1.39 | 5.9E-10 | 0.75 | 8.3E-07 | -0.64 | 8.8E-05 | |
| YHR195W (*NVJ1*) | -2.39 | 1.7E-06 | 0.39 | 0.4 | -2.00 | 1.7E-04 | |
| YIL165C | -3.22 | < 2e-16 | -0.29 | 0.2 | -3.51 | 1.1E-14 | (2) |
| YIL164C (*NIT1*) | -3.76 | 5.5E-07 | -0.37 | 0.9 | -4.13 | 3.0E-06 | (2) |
| YJL213W | -3.41 | 9.4E-10 | 1.26 | 2.4E-05 | -2.15 | 1.2E-04 | |
| YJL132W | -1.37 | 3.7E-07 | 0.67 | 0.03 | -0.70 | 3.1E-03 | |
| YJR015W | 4.16 | 4.2E-13 | 1.19 | 3.2E-05 | 5.35 | 1.5E-32 | (3) |
| YJR072C (*NPA3*) | -2.14 | < 2e-16 | 1.46 | 3.4E-26 | -0.68 | 1.6E-07 | |
| YKL163W (*PIR3*) | -1.19 | 1.6E-06 | -0.07 | 0.6 | -1.26 | 9.8E-08 | |
| YKL095W (*YJU2*) | -1.55 | 2.1E-06 | 0.60 | 0.03 | -0.95 | 1.9E-04 | |
| YKL012W (*PRP40*) | -1.47 | 9.6E-06 | 0.24 | 0.9 | -1.24 | 1.1E-04 | |
| YLL022C (*HIF1*) | 1.02 | 7.2E-06 | -0.64 | 6.6E-03 | 0.38 | 0.04 | |
| YLL007C (*LMO1*) | 1.82 | 5.9E-08 | -1.40 | 3.3E-08 | 0.42 | 0.03 | |

| | | | | | | |
|---|---|---|---|---|---|---|
| YLR375W (*STP3*) | -1.26 | 7.2E-06 | 0.47 | 0.03 | -0.78 | 1.6E-02 |
| YML048W (*GSF3*) | -1.06 | 4.1E-08 | 1.13 | 4.7E-17 | 0.06 | 0.4 |
| YMR091C (*NPL6*) | 1.16 | 2.5E-08 | -1.26 | 8.4E-14 | -0.10 | 0.6 |
| YOR304W (*ISW2*) | 1.34 | 5.7E-12 | -0.89 | 2.4E-08 | 0.45 | 6.1E-03 |

For consistency with the figures, the fold changes are log2-transformed so that zero indicates no difference and one indicates a two-fold difference. Positive values indicate higher abundance in BY compared to RM. FP: footprints. NS: neither mRNA nor footprint difference was significant. (1) "Dubious" ORF, footprint data shows translated region only partially overlaps with annotation. The TE difference is due to a nonsense SNP in BY that results in early termination compared to RM. (2) YIL165C is a "dubious" ORF immediately downstream of YIL164C (*NIT1*); in RM, these two ORFs form a single, consistently translated ORF (Figure 7C). (3) Putative protein with frameshift in RM that leads to premature termination. Note that "dubious" ORFs were not included in our analyses of nonsense SNPs so that YDR133C and YJR015W were not included in those analyses.

Table 3 – Effects of translation in genes with significant TE

| Significant difference | Direction of differences | Magnitude of differences | Parent | Hybrid |
|---|---|---|---|---|
| Footprint only | – | – | 611 (27%) | 27 (25%) |
| mRNA and footprint | same | Footprint > mRNA | 690 (31%) | 10 (9%) |
| mRNA and footprint | same | mRNA > footprint | 229 (10%) | 4 (4%) |
| mRNA only | – | – | 420 (19%) | 37 (35%) |
| mRNA and footprint | opposite | – | 159 (7%) | 5 (5%) |
| neither | – | – | 119 (5%) | 23 (22%) |
| Sum | – | – | 2,228 | 106 |

Numbers shown in this table are based on a Bonferroni-corrected significance threshold.

Table 4 – Tests for directional preferences in the effects of translational differences in different data sets

| Compared groups of TE genes | BY / RM (Bonferroni) | | BY / RM (FDR) | | Scer / Spar (Artieri & Fraser) | | Scer / Spar (McManus et al.) | |
|---|---|---|---|---|---|---|---|---|
| | Parents | Hybrid | Parents | Hybrid | Parents | Hybrid | Parents | Hybrid |
| Reinforced vs. buffered | 1.1 (0.3) | 0.7 (0.09) | 1.3 (8e-10) | 0.2 (1e-18) | 1.6 (3e-13) | 1.0 (0.7) | 0.3 (3e-145) | 0.2 (3e-118) |
| Reinforced vs. buffered & inverted | 0.9 (0.02) | 0.6 (0.03) | 0.8 (3e-13) | 0.2 (1e-22) | 0.9 (0.2) | 0.6 (9e-10) | 0.3 (4e-190) | 0.2 (1e-138) |
| Reinforced & FP only vs. buffered | 2.0 (2e-49) | 0.9 (0.7) | 1.6 (2e-45) | 1.4 (0.001) | 1.6 (3e-15) | 1.4 (3e-6) | 0.6 (8e-52) | 0.7 (9e-19) |
| Reinforced & FP only vs. buffered & inverted | 1.6 (7e-27) | 0.8 (0.3) | 1.0 (0.4) | 1.2 (0.03) | 1.0 (0.4) | 0.9 (0.2) | 0.5 (7e-82) | 0.6 (2e-28) |

Each cell shows the ratio of TE genes where translation increases a gene expression difference (with or without "FP only" TE genes) vs. genes where translation opposes the gene expression difference (with or without inverted genes). Values greater than one indicate more genes where translation increases the gene expression difference, and values less than one indicate more genes with opposing effects. In parentheses are the p-values of a chi-squared test of the hypothesis that the two compared groups are observed at the same frequency.

Figure Legends

Figure 1 – Global mRNA and footprint abundance

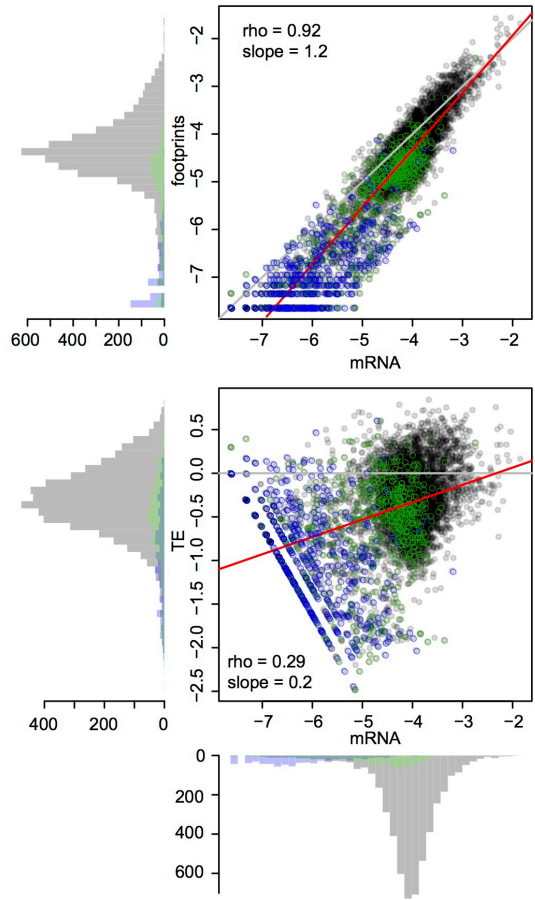

Shown are log10-transformed normalized read counts in the BY strain. Top panel: mRNA vs. footprint abundance. The red line shows the regression of footprint on mRNA abundance. The grey line indicates identity. Bottom panel: TE as a function of mRNA abundance. The grey line denotes identity between footprint and mRNA levels (i.e. log10(TE) = 0). The red line shows the regression of TE on mRNA abundance. Throughout the figure, transparent grey points are "verified" ORFs, green points are "uncharacterized" ORFs and blue points are "dubious" ORFs.

Figure 2 – Expression in BY vs. RM and ASE in the hybrid

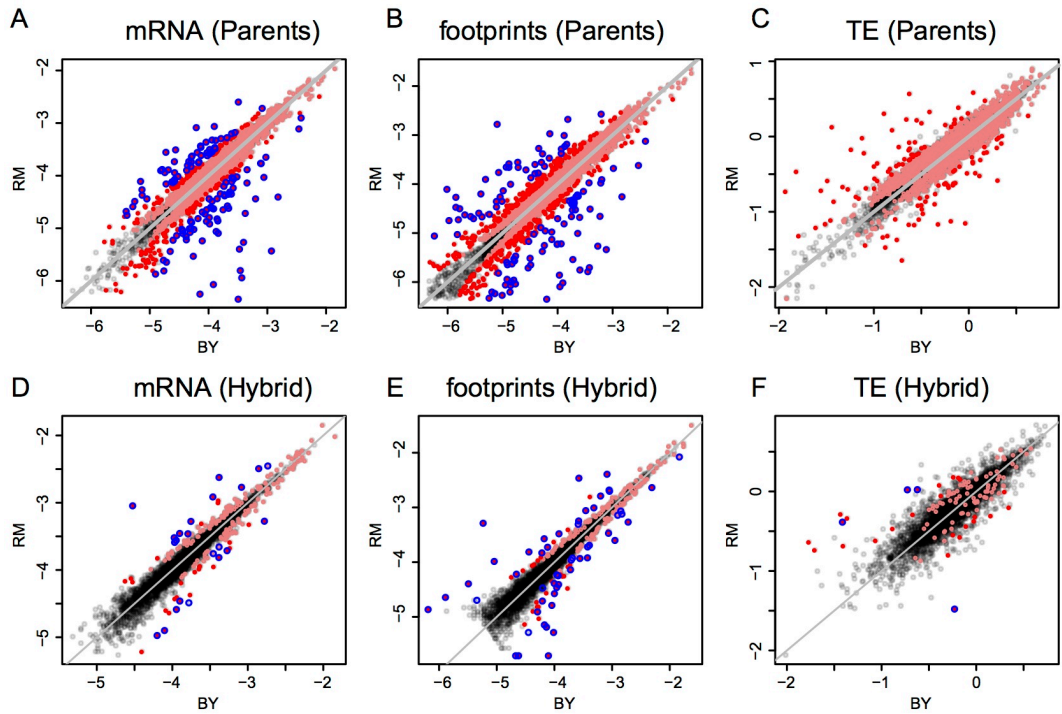

Shown are log10-transformed normalized read counts based on the downsampled data (Methods). Grey diagonal lines mark identity. Light red points are genes with significant differences (Bonferroni corrected: p < 9e-6 in parent data and p < 1.5 e-5 in hybrid data), darker red points are significant genes with a fold change ≥ 2. The blue circles denote genes that were called significant by DESeq (Benjamini-Hochberg adjusted p < 0.05). A-C: parental comparisons, D-F: hybrid ASE. Note that in F) only four of the nine genes with significant TE difference identified by DESeq are shown, the remaining five had abundance too low to be included in the downsampled data (Methods).

Figure 3 – mRNA vs. footprint differences

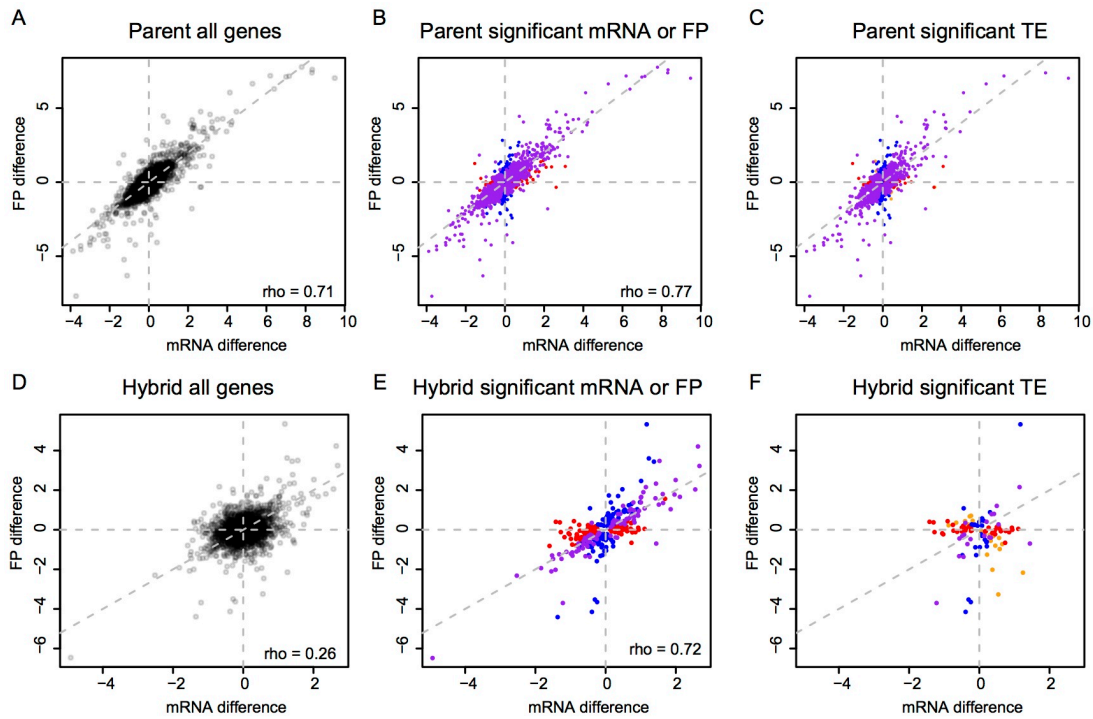

Shown are log2-transformed fold changes. A-C: parents, D-F: hybrid ASE. Grey dashed lines are the diagonal. Left column: all genes. Middle column: genes with a significant (Bonferroni corrected: p < 9e-6 in parent data and p < 1.5 e-5 in hybrid data) mRNA (red), footprint (blue) or both mRNA and footprint (purple) difference. Right column: genes with a significant TE difference. Red: genes with only a significant mRNA difference, blue: genes with only a significant footprint difference, purple: genes with both a significant mRNA and footprint difference, orange: genes with neither a significant mRNA nor a significant footprint difference.

Figure 4 – *Cis* and *trans* effects

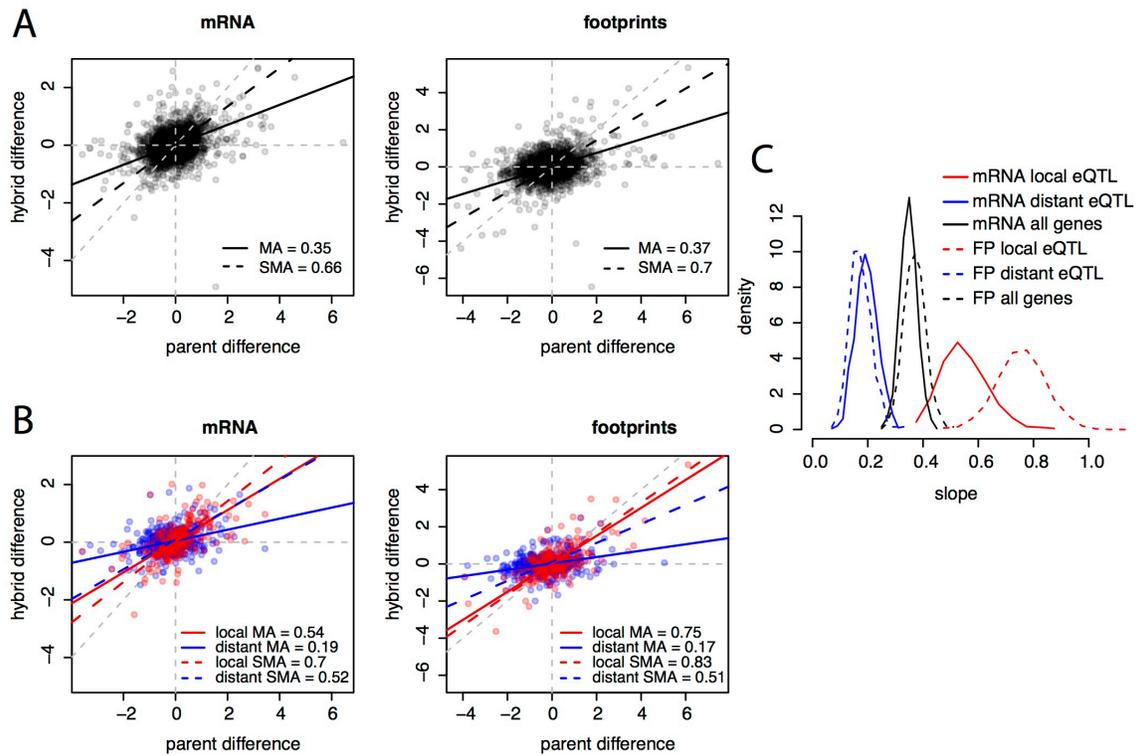

A. Parental differences (estimated based on SNP allele counts) on the x-axes, and hybrid differences on the y-axes, for all genes. Black lines show the slope of the relationship between hybrid and parental differences. The legends indicate the values of these slopes. MA: major axis estimate; SMA: standardized major axis estimate. B. as in A), but only for genes with eQTL in [29]. Red: genes with a local but no distant eQTL, blue: genes with a distant but no local eQTL, purple: genes with both a local and a distant eQTL. Colored lines show the respective regressions of hybrid on parental differences. C. bootstrapped distributions of MA slope estimates. Results from SMA were qualitatively similar.

Figure 5 – Relationship between mRNA differences and footprint differences within and between species

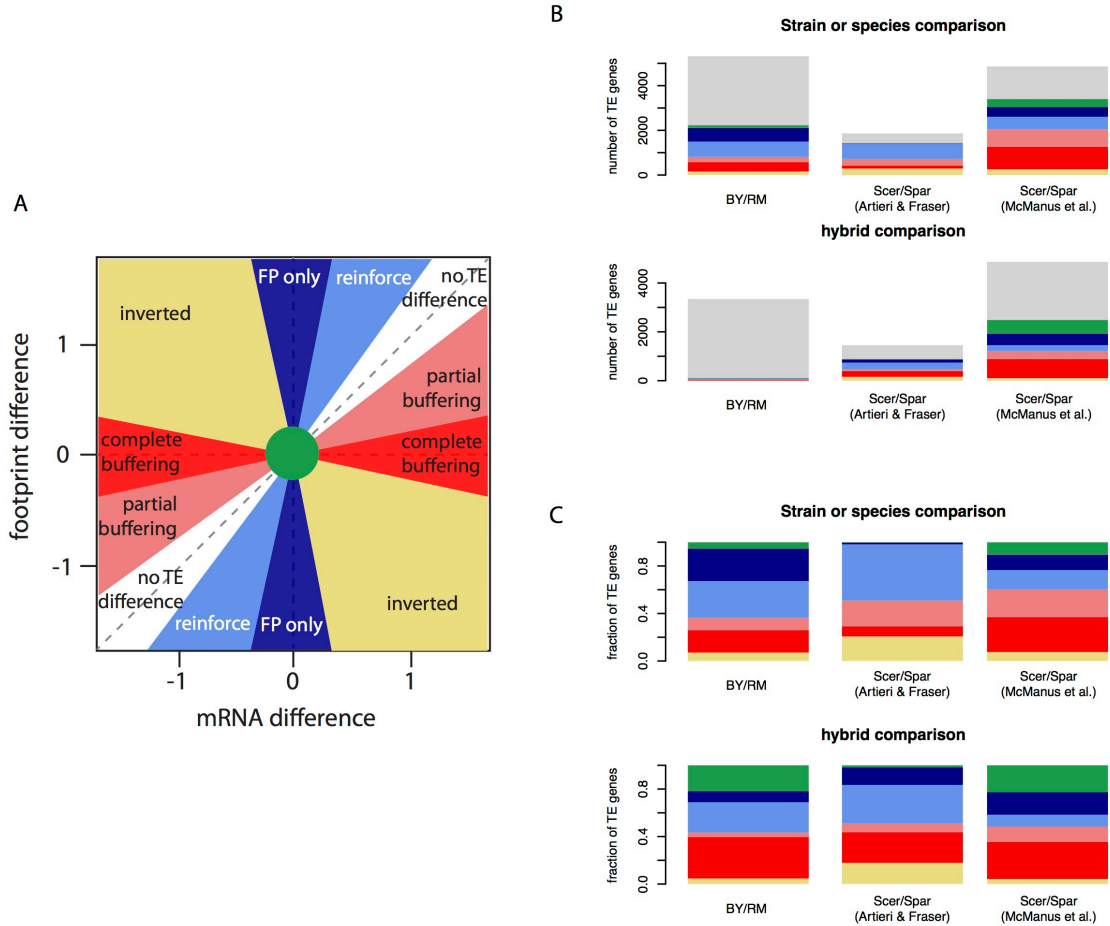

A. Schematic representation of the possible relationships between mRNA differences and footprint differences. B. Observed distribution of all analyzed genes in three data sets. The color scheme is the same is in A), with light grey indicating genes without a TE difference. For BY / RM, significance was determined using a Bonferroni-corrected p-value of < 0.05. *Scer*: *Saccharomyces cerevisiae*. *Spar*: *Saccharomyces paradoxus*. The interspecies data were analyzed from published datasets ([35] and [34]). C) As in B), but showing the fraction of genes with a certain relationship among genes with a significant TE difference.

Figure 6 – Comparison of mRNA and footprint differences to pQTL effects

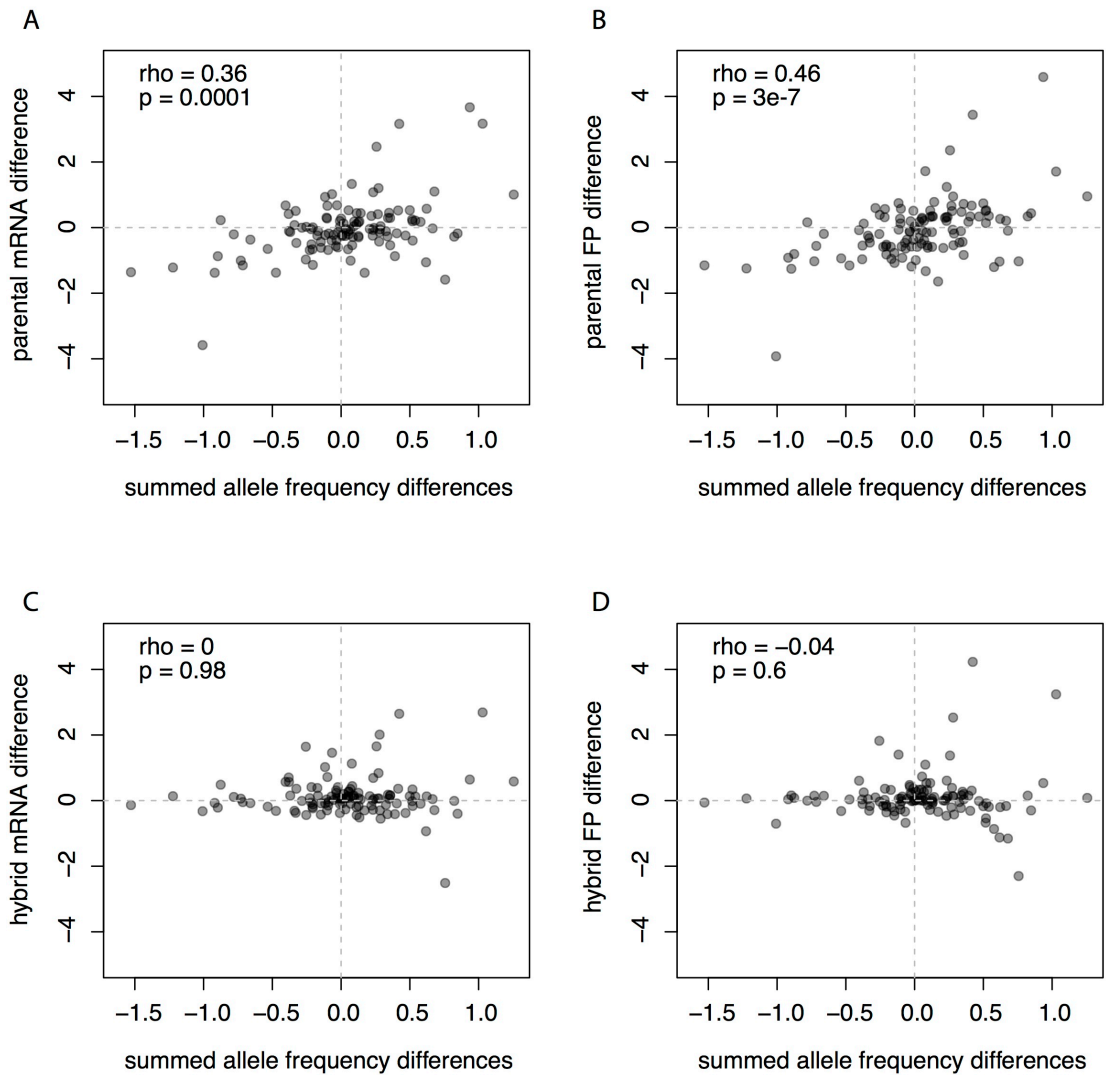

For each gene, the pQTL effects are shown as the sum of allele frequency differences at all pQTL identified by a bulk segregant approach [8]. The pQTL effect sizes shown on the x axis are identical in all four panels and are compared to A) parental mRNA differences, B) parental footprint differences, C) hybrid allele-specific mRNA differences and D) hybrid allele-specific footprint differences.

Figure 7 – Examples of patterns of translation at putative premature stop codons

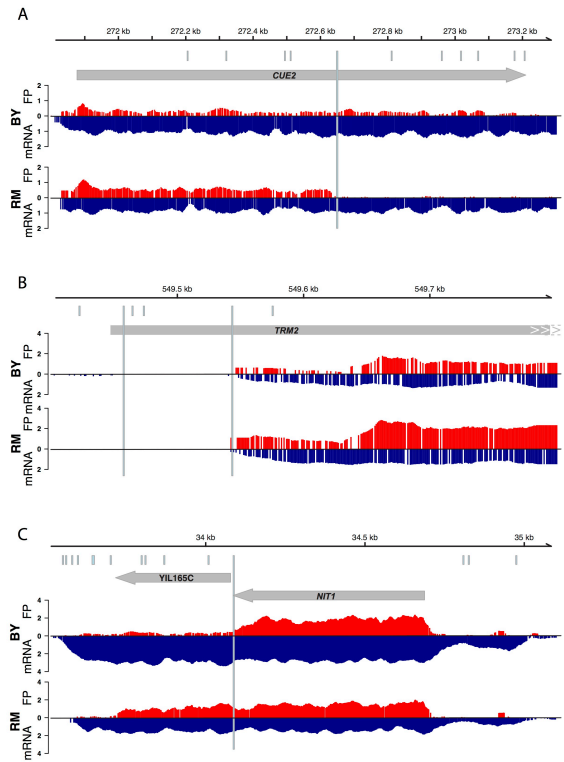

Grey arrows indicate the position and strand of ORFs. Footprints (red) and mRNA (blue, inverted scale) for BY and RM are plotted beneath. The positions of putative premature stop codons in BY or RM are shown as light blue, longer horizontal bars, while all sequence differences between BY and RM are shown as light blue tickmarks above the ORF. The mRNA and footprint densities are shown as log transformed numbers of read starts in 30 bp wide smoothed windows. They are only shown for the strand of the displayed ORFs. A. An example of a premature translation termination in *CUE2* in RM compared to BY. B. Two putative nonsense SNPs in *TRM2* are in fact upstream of the translated and transcribed ORF. C. The gene *NIT1* in BY is the result of a premature termination of a full length ORF that in RM includes the downstream ORF YIL165C.

# Supplementary Figure Legends

Supplementary Figure S1 – Comparison to Ingolia *et al.* 2009 data

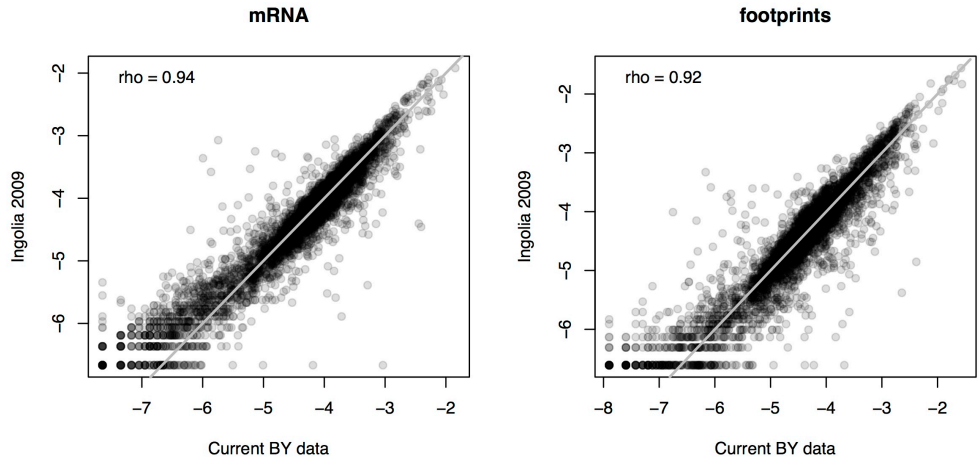

Shown are log10 transformed normalized read counts. The grey line marks identity.

Supplementary Figure S2 – mRNA vs. footprint differences identified by DESeq

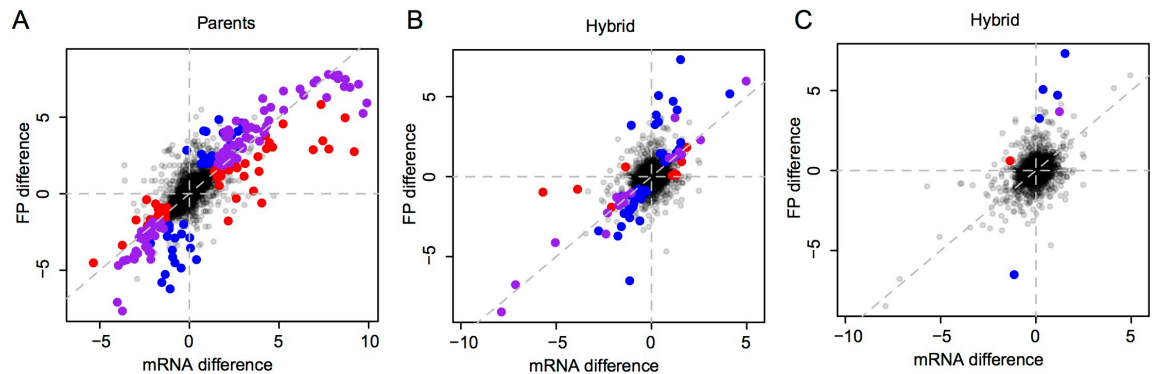

Shown are log2-transformed fold changes. A: parents, B & C: hybrid ASE. A & B: genes with a significant (Benjamini-Hochberg corrected $p < 0.05$) mRNA (red), footprint (blue) or both mRNA and footprint (purple) difference. C: genes with a significant TE difference. Red: genes with only a significant mRNA difference, blue: genes with only a significant footprint difference, purple: genes with both a significant mRNA and footprint difference.

Supplementary Figure S3 – Reproducibility of hybrid measurements

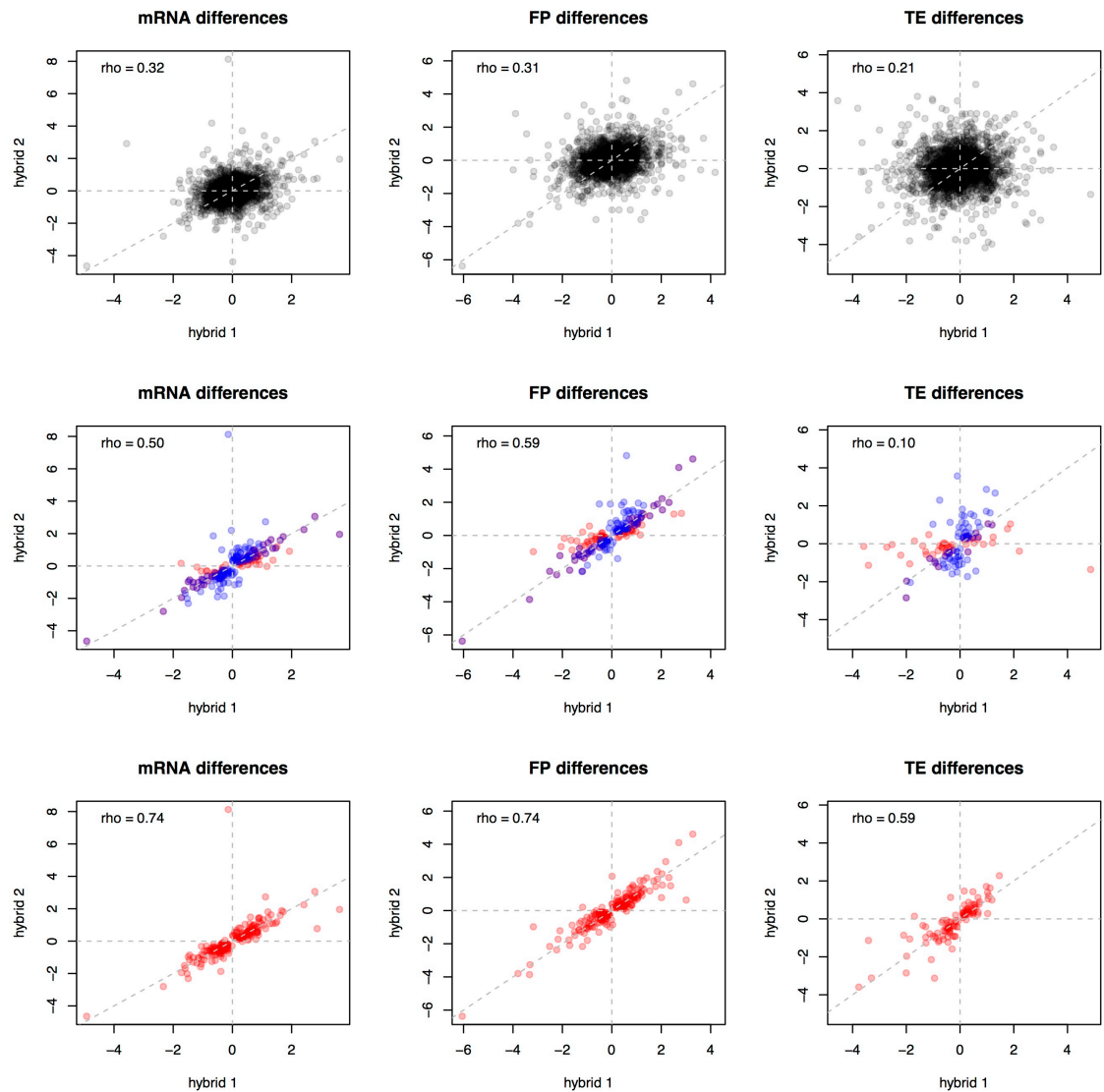

Shown are log2-transformed fold changes. Grey diagonals mark identity. Top row: all genes. Middle row: significant genes in one and / or the other replicate. Bottom row: significant genes in the combined hybrid data. Spearman correlation coefficients between replicates are given in each panel.

Supplementary Figure S4 – Spurious correlations induced by correlations between a log ratio and its denominator

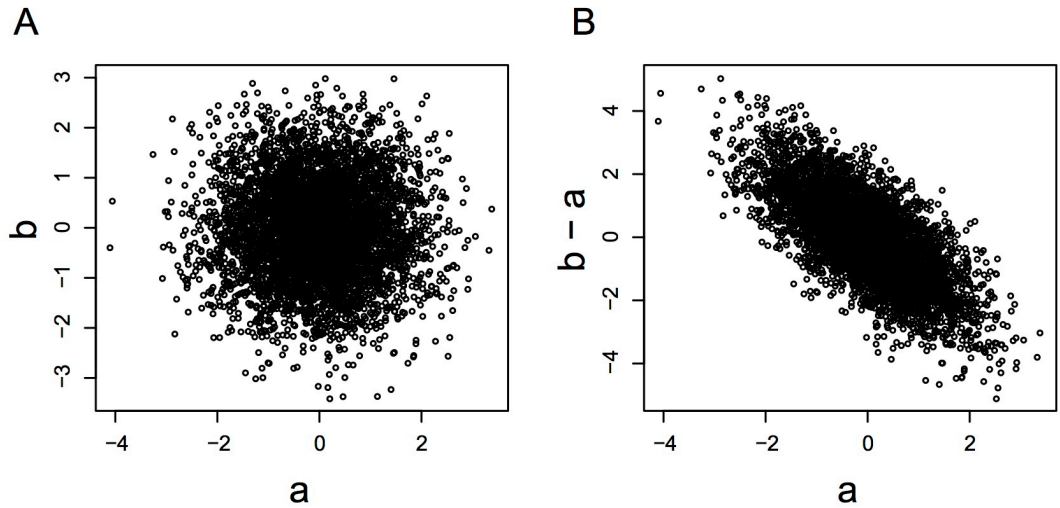

A. A scatterplot of two random samples *a* and *b* of size 5,000 from a standard normal distribution with mean = 0 and standard deviation = 1. Note that *a* and *b* are entirely uncorrelated. B. The correlation between the quantity *b* - *a* and *a* is negative and highly significant because of regression to the mean. For example, when *a* happens to be large by chance, the corresponding value of *b* will usually be closer to the mean than *a* because it is unlikely that a large value is sampled two times by chance. Therefore, the quantity *b* - *a* is systematically more likely to be less than zero for *a* > 0. If *a* and *b* are interpreted as the logarithms of mRNA and footprint differences, *b* - *a* is equivalent to the corresponding TE differences. A negative correlation between TE differences and mRNA differences is thus not by itself sufficient to infer translational buffering.

Supplementary Figure S5 – Replicate noise in different hybrid datasets

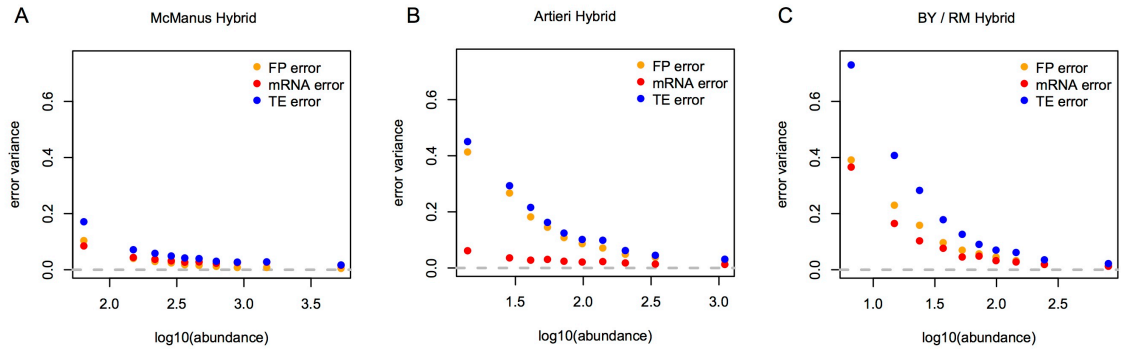

For each gene in each dataset, we calculated the log2 fold change between the alleles in the hybrid for mRNA, footprint, and TE separately for each of the two replicate datasets. The genes in each dataset were divided into 10 bins of increasing mRNA expression level. Within each bin, the average variance in the log2 fold change that is due to measurement error was calculated using the meas.est() function in the R smatr package [33] and plotted as a function of the mean abundance of the genes in the given bin. A & B: data from the published interspecies hybrid comparisons in McManus et al. [34] (A) and Artieri & Fraser [35] (B). C: Data from the BY / RM hybrid. In all datatsets, error is higher for genes with lower abundance. Footprints typically have higher error than mRNA. However, the degree to which these two data types differ varies between datasets (e.g. compare A to B). The error variance in TE is the sum of the errors in mRNA and footprints.

# Supplementary Note 1 – Genomic sources of mRNA and footprint fragments

We generated 189 million (M) and 222 M ribosomal footprint reads from BY and RM, respectively (see Supplementary Table S1 for read and alignment statistics for all samples). In parallel, we gathered mRNA data from the same yeast cultures. Of the reads, 82 M (43%) for BY and 151 M (68%) for RM mapped to unique positions in the genome. The difference in the percentage of uniquely mapping reads is likely due to differences in the efficacy of ribosomal RNA (rRNA) depletion during library construction. rRNA is transcribed from a highly repetitive region of the genome (s. below) and greatly outnumbers mRNA in exponentially growing yeast cells [1] so that even minor differences in the efficacy of rRNA depletion can lead to large differences in rRNA retention. As a consequence, more uniquely mapping mRNA sequencing reads are available in samples with more effective rRNA depletion.

We determined which genomic features were the source of the footprint and mRNA reads. These analyses were conducted on the data from the BY strain, in order to allow us to directly use the reference genome annotation (Supplementary Table S2). Of uniquely mapping mRNA reads, the vast majority (84%) corresponded to either protein coding sequences (CDS) or untranslated regions (UTRs; 17%, the sum can be more than 100% because of overlapping annotations). For footprints, 97% of uniquely mapping reads aligned to CDS, while 6% mapped to UTRs. The higher fraction mapping to CDS in footprints fits the expectation that translating ribosomes should be preferentially found on coding regions, rather than on UTRs. Notably, of mRNA reads that uniquely mapped to UTRs, 35% mapped to 5' UTRs, while this fraction was nearly twice as high (68%) in the footprints. A higher density of ribosomes in 5'UTRs than expected based on mRNA abundance may be due to ribosomes that translate upstream open reading frames (uORFs) on 5'UTRs but not 3'UTRs [2,3]. Of the reads that mapped to multiple locations in the genome, many fewer mapped to CDS and UTRs (Supplementary Table S2). These "repeat reads" were heavily dominated by the ribosomal rRNA genes: 90% of mRNA and 85% of footprint reads mapped there. The rRNA genes are each represented by a small number of gene annotations in the yeast reference genome sequence (two annotations for 35S pre-rRNA and six annotations for different variants of 5S rRNA) which represent 100-200 tandem repeats of the rDNA locus [4]. The large number of rRNA reads therefore reflects rRNA transcription across all rDNA repeats and is in line with the high amounts of rRNAs in exponentially growing yeast [1]. In this work, we considered only reads with unique alignments.

References for Supplementary Note S1

# Supplementary Note 2 – Analyses based on the DESeq framework

The DESeq analysis framework employs a negative binomial distribution to identify genes with significant expression differences while explicitly accounting for differences in the total number of reads in each sample and differences in expression noise for genes with different expression levels [1].

In the parent strains, DESeq identified 189 genes with significant (5% FDR) mRNA differences, and 145 significant footprint differences (Table 1). These numbers are substantially lower than those obtained using the binomial test due to that fact that in data without replicates, DESeq estimates noise from the differences between the two samples. Because true strain differences are conflated with technical and biological variation, this procedure overestimates the noise, resulting in a conservative test. Nevertheless, the agreement between the mRNA and footprint differences identified by DESeq was very high: of the 189 genes with an mRNA difference, 101 also had a significant footprint difference (Fisher's exact test $p < 2.2e-16$, odds ratio = 157). The direction of effect agreed for 97% (227 / 233) genes with either a significant mRNA and / or a significant footprint difference (Supplementary Figure S2A). It is not possible to calculate significant differences in TE using DESeq without biological replicates.

In the hybrid data, DESeq called 40 genes as having significant (5% FDR) allele-biased mRNA expression and 70 genes with significant allele-biased footprint counts. These numbers are again conservative compared to the binomial test. However, the difference in the numbers of significant genes between DESeq and the binomial test is smaller for the hybrid than the parental comparison. This is because for the hybrid data, DESeq uses the two biological replicates, which adds power to the test. Again, the mRNA and footprint differences in the hybrid were very similar: of the 40 genes with an mRNA difference, 26 also had a footprint difference (FET $p < 2.2e-16$, odds ratio = 166). The direction of allelic expression bias agreed for 98% (82 / 84) of genes with an mRNA and / or a footprint difference (Supplementary Figure S2B).

DESeq identified 9 genes with significant TE differences (Supplementary Table S4 & Supplementary Figure S2B). Four of these were also found by the binomial test, and the remaining five genes had been excluded from analysis with the binomial test due to low expression. The five genes only identified by DESeq contained two additional cases where the TE difference is due to a longer ORF in RM is broken up into two annotated ORFs in BY (Supplementary Table S4).

We portioned these 9 TE genes to ask if reinforcing or buffering interactions predominate (Supplementary Table S5). There were 6 genes with a footprint but no mRNA difference, 1 gene with only an mRNA difference, 1 gene with both an mRNA and a footprint difference where the footprint difference was larger than the mRNA difference, and one gene where neither mRNA nor footprints were significantly different between alleles. Thus, there is one TE gene consistent with buffering and 7 consistent with footprints reinforcing or generating an expression difference ($\chi^2 = 4.5$, $p = 0.03$).

In sum, DESeq identified substantially fewer genes with significant differences, as expected for a conservative method [2] in an experiment with few replicates. However, the important patterns presented in the main text held using these much smaller sets of genes. Significant mRNA and footprint differences agreed very well in both parents and the hybrid. Further, among the few genes flagged as having significant differences in TE in the hybrid, there was an excess of genes with larger footprint than mRNA differences.

## References for Supplementary Note S2

# Supplementary Note S3 – Directional effects of translation differences within and between yeast species

To test whether there is a directional excess of translational effects, we divided up the TE genes (i.e. genes where the footprint difference was significantly different from the mRNA difference) as shown in Figure 5 and Table 3. There are several choices that can be made in making these comparisons.

For example, the authors of the two comparisons of the *S. cerevisiae* and *S. paradoxus* yeast species [1,2] examined the subset of TE genes that also had an mRNA difference. Within this subset, they compared the number of reinforced genes to the sum of the buffered, completely buffered and inverted genes, i.e. all cases where the translation difference opposes the mRNA difference ("opposing" genes). These analyses leave out "FP only" genes, although "FP only" genes and reinforced genes both increase the footprint difference relative to the mRNA difference (we call these two groups "increasing" genes). Further, it is not necessarily obvious that inverted genes should be grouped with the two "buffered" categories, because the absolute magnitude of the resulting difference in protein synthesis need not be smaller than the mRNA difference. Indeed, McManus et al. only included inverted genes in their set of opposing genes if the absolute footprint difference was smaller than the absolute mRNA difference [1].

We systematically conducted all possible comparisons between different definitions of opposing and increasing genes (Figure 5 & Tables 3 & 4). Between the BY and RM parents there were more TE genes where translation increased rather than decreased or inverted the footprint difference relative to the mRNA difference, but this effect was dependent on the inclusion of the "FP only" genes in the "increased" genes. The result further depended on the precise significance cutoff used to group the TE genes (Table 4 & Supplementary Table S5).

In the BY / RM hybrid data, there were more TE genes where translation increased the footprint difference relative to the mRNA difference, but only when the "FP only" genes were included in the analyses, and only when ASE TE genes were defined using the more liberal FDR criterion (Table 4).

Given that the BY / RM result depended on the precise way in which the TE genes are grouped and compared, we conduced the same analyses for the two interspecies comparisons [1,2]. As shown in the main text, the results from [1] were robust to the precise comparison and always showed an excess of opposing effects of translation in both the parent and the hybrid comparison.

In the hybrid data from [2], we saw the same effect when conducting the comparison exactly as described by the authors (s. above). However, when "FP only" genes are included in the comparison and inverted genes are excluded, the remaining opposing genes are in the minority compared to the increasing genes (Table 4). The other two possible comparisons show no significant preference for or against the respective sets of opposing or increasing genes. In the parent data from [2], there is a significant excess of increasing genes when inverted genes are excluded from the opposing genes. When the inverted genes are included, the number of increasing and opposing effects are not statistically different (Table 4).

The two published reports of predominant opposing effects of translation between yeast species were respectively based on additional analyses. McManus et al. reported that TE differences showed a negative correlation with mRNA differences [1]. The same pattern is visible (although it was not highlighted as such) in the data by Artieri & Fraser (Figure 2A top panel in [2]) as well as in our own data (not shown). However, the TE difference is the ratio of the footprint difference and the mRNA difference. Comparisons between ratios and their components can induce "spurious" correlations [3,4] (Supplementary Figure S4). A negative correlation between TE differences and mRNA differences therefore does not provide evidence for translational buffering by itself. An excess of opposing effects in Artieri & Fraser [2] was further supported by the observation that the slope of the regression of footprint differences on mRNA differences was less than one. An alternative explanation for this slope estimate may be regression to the mean ([5] p. 58). In the presence of measurement noise, and when two observations are on similar scales (as is the case for mRNA and footprint differences), regression slopes are less than one.

# References for Supplementary Note S3

## Supplementary Table Legends

Supplementary Table S1 – Sequencing and alignment statistics

| Strain | Data type | Raw reads | Parent comparison: unique alignments | | ASE analyses: unique & no mismatch | | ASE analyses: Unique, no mismatch & spans a SNP | |
|---|---|---|---|---|---|---|---|---|
| Reference genome | | | BY | Edited BY[1] | BY | RM | BY | RM |
| BY parent | Footprint | 189 | 82 | - | 74 | - | 3.7 | - |
| BY parent | mRNA | 146 | 53 | - | 46 | - | 2.4 | - |
| RM parent | Footprint | 222 | - | 151 | - | 129 | - | 6.7 |
| RM parent | mRNA | 129 | - | 52 | - | 46 | - | 2.5 |
| BY/RM diploid 1 | Footprint | 103 | - | - | 33 | 32 | 0.9 | 0.9 |
| BY/RM diploid 1 | mRNA | 98 | - | - | 27 | 26 | 0.7 | 0.7 |
| BY/RM diploid 2 | Footprint | 108 | - | - | 44 | 42 | 1.2 | 1.2 |
| BY/RM diploid 2 | mRNA | 113 | - | - | 28 | 28 | 0.8 | 0.8 |

Numbers are given in millions of reads. [1]A version of the BY reference genome with all known single nucleotide differences set to the RM allele.

Supplementary Table S2 – Genomic sources of mRNA and footprint reads in the BY parent

|  | mRNA | | Ribosomal footprints | |
|---|---|---|---|---|
|  | unique | repetitive | unique | repetitive |
| CDS | 44.4M (84%) | 5.8M (6.6%) | 79.2M (97%) | 12.2M (12%) |
| UTRs[1] | 8.9M (17%) | 66k (0.1%) | 4.7M (5.7%) | 145k (0.1%) |
|     5'UTR | 3.1M (35%)[1] | 34k (52%)[1] | 3.2M (68%)[1] | 69k (48%)[1] |
|     3'UTR | 5.9M (66%)[1] | 32k (48%)[1] | 1.5M (32%)[1] | 75k (52%)[1] |
| rRNA | 246k (0.5%) | 79M (90%) | 950k (1.2%) | 88M (85%) |
| tRNA | 73k (0.1%) | 466k (0.5%) | 228k (0.3%) | 1.2M (1.2%) |
| Other noncoding: snoRNA, snRNA, ncRNA | 468k (0.9%) | 727 | 217k (0.3%) | 264 |
| Total | 53M | 88M | 82M | 103M |

Percentages can sum to more than 100 due to overlapping annotations

[1]percent of all UTRs

Supplementary Table S3 – FDR-based differential expression statistics

| Comparison | Reads | Data | Analyzed genes | 2-fold | Binomial test[1] | Intersect |
|---|---|---|---|---|---|---|
| Parent | All | mRNA | 5,316 | 331 | 4,575 | 331 (6%) |
| Parent | All | Footprint | 5,316 | 514 | 4,669 | 512 (10%) |
| Parent | All | TE | 5,316 | 135 | 4,256 | 135 (3%) |
| Parent | SNP | mRNA | 3,342 | 249 | 1,159 | 225 (7%) |
| Parent | SNP | Footprint | 3,342 | 475 | 1,486 | 441 (13%) |
| Parent | SNP | TE | 3,342 | 329 | 1,155 | 278 (8%) |
| Hybrid | SNP | mRNA | 3,342 | 100 | 529 | 65 (2%) |
| Hybrid | SNP | Footprint | 3,342 | 194 | 617 | 128 (4%) |
| Hybrid | SNP | TE | 3,342 | 216 | 638 | 148 (4%) |

[1] q-value < 0.05

Supplementary Table S4 – Strong *cis* effects on translation identified by DESeq

|  | TE | mRNA |  | FP |  | Identified by binomial test? | Notes |
|---|---|---|---|---|---|---|---|
|  | p-value | Log2(fold change) | p-value | log2(fold change) | p-value |  |  |
| YBR012C | 2.6E-06 | 1.12 | 1.4E-03 | 4.69 | 3.0E-07 |  |  |
| YDL231C (*BRE4*) | 1.6E-05 | 0.16 | 0.6 | 3.23 | 3.1E-06 | Yes |  |
| YDR133C | 9.9E-07 | 1.22 | 1.8E-06 | 3.66 | 1.8E-15 | Yes | (1) |
| YJL108C (*PRM10*) | 3.7E-05 | 0.35 | 0.5 | 5.05 | 1.2E-05 |  | (2) |
| YJR015W | 1.2E-11 | -1.16 | 3.1E-03 | -6.53 | 5.1E-23 | Yes | (3) |
| YJR072C (*NPA3*) | 2.3E-08 | -1.38 | 1.9E-06 | 0.59 | 0.03 | Yes |  |
| YNL020C (*ARK1*) | 7.5E-11 | -0.69 | 0.1 | -Inf[1] | 1.5E-15 |  |  |
| YNR065C | 8.3E-05 | 0.92 | 0.06 | Inf[1] | 2.3E-03 |  | (4) |
| YPR192W (*AQI1*) | 2.9E-06 | 1.51 | 0.2 | 7.30 | 6.1E-23 |  |  |

[1]Infinite fold changes indicate that there were zero counts in one of the groups. Such genes were excluded from the binomial tests reported in the main text. Genes not identified by the binomial test all had counts below the inclusion criteria for binomial testing. (1) "Dubious" ORF, footprint data shows translated region only partially overlaps with annotation. The TE difference is due to a nonsense SNP in BY that results in early termination compared to RM. (2) Based on the parental read data, YJL108C forms one ORF in RM with its upstream neighbor YJL107C. The combined ORF in RM is interrupted by a stop mutation in BY, resulting in two separate gene annotations. (3) Putative protein with frameshift in RM that leads to premature termination. Note that "dubious" ORFs were not included in our analyses of nonsense SNPs so that YDR133C and YJR015W were not included in those analyses. (4) Similar to (1), and uncharacterized ORF that in RM forms one ORF with the upstream "uncharacterized" YNR066C.

Supplementary Table S5 – Effects of translation in TE genes derived from alternative significance criteria

| Significant difference | Direction of differences | Magnitude of differences | Parent FDR | Hybrid FDR | Hybrid DESeq |
|---|---|---|---|---|---|
| mRNA and footprint | same | Footprint > mRNA | 1,638 (38%) | 40 (6%) | 1 |
| Footprint only | – | – | 497 (12%) | 193 (30%) | 6 |
| mRNA and footprint | same | mRNA > footprint | 904 (21%) | 20 (3%) | 0 |
| mRNA only | – | – | 401 (9%) | 147 (23%) | 1 |
| mRNA and footprint | opposite | – | 778 (18%) | 21 (3%) | 0 |
| neither | – | – | 38 (1%) | 217 (34%) | 1 |
| Sum | – | – | 4,256 | 638 | 9 |

FDR: genes with q-values < 0.05 in the binomial tests.

Supplementary Table S6 – Effects of translation in TE genes in published interspecies comparisons

| Significant difference | Direction of differences | Magnitude of differences | McManus Parent | McManus Hybrid | Artieri Parent | Artieri Hybrid |
|---|---|---|---|---|---|---|
| Footprint only | – | – | 443 | 471 | 22 | 132 |
| mRNA and footprint | same | Footprint > mRNA | 552 | 249 | 669 | 287 |
| mRNA and footprint | same | mRNA > footprint | 794 | 319 | 307 | 66 |
| mRNA only | – | – | 1,001 | 778 | 120 | 229 |
| mRNA and footprint | opposite | – | 258 | 108 | 293 | 159 |
| neither | – | – | 357 | 567 | 4 | 15 |
| Sum | – | – | 3,405 | 2,492 | 1,415 | 888 |

## Supplementary Data Legends

Supplementary Data S1 – Influence of annotated nonsense and frameshift mutations on translation

Supplementary Data S2 – Complete raw and processed data